# Experimental demonstration of correlated flux scaling in photoconductivity and photoluminescence of lead-halide perovskites.

**Nov. 09, 2017**

Hee Taek Yi[1], Pavel Irkhin[1,†], Prakriti P. Joshi[2], Yuri N. Gartstein[3], Xiaoyang Zhu[2] and Vitaly Podzorov[1,*]

[1] Dept. of Physics, Rutgers University, Piscataway, NJ 08854, USA;
[2] Dept. of Chemistry, Columbia University, New York, NY 10027, USA;
[3] Dept. of Physics, University of Texas at Dallas, Richardson, TX 75080, USA.

\* E-mail:  podzorov@physics.rutgers.edu

**Lead-halide perovskites attracted attention as materials for high-efficiency solar cells and light emitting applications. Among their attributes are solution processability, high absorbance in the visible spectral range and defect tolerance, as manifested in long photocarrier lifetimes and diffusion lengths. The microscopic origin of photophysical properties of perovskites is, however, still unclear and under debate. Here, we have observed an interesting universal scaling behavior in a series of (hybrid and all-inorganic) perovskite single crystals investigated via simultaneous measurements of the Hall effect, photoconductivity and photoluminescence. A clear correlation between photoconductivity and photoluminescence as functions of the incident photon flux is observed. While photoconductivity exhibits a crossover in the power-law dependence between power exponents 1 and 1/2, photoluminescence exhibits a crossover between power exponents 2 and 3/2. This correlation is found in all the studied compounds irrespective of the cation type (organic or inorganic) or crystallographic phases. We propose phenomenological microscopic mechanisms that explain these interesting non-trivial power exponents and crossovers between them in this broad class of lead-halide perovskites.**

---

[†] Current address: Micron.Inc, Boise, Idaho, USA.

**INTRODUCTION**

Hybrid lead-halide perovskites, the materials composed of an inorganic sublattice of corner-sharing lead-halide octahedra (e.g., $PbI_6^-$ or $PbBr_6^-$) and an organic sublattice of methyl-ammonium $CH_3NH_3^+$ ($MA^+$) or formamidinium $CH_2N_2H_4^+$ ($FA^+$) molecular cations, form the materials basis for the emergent perovskite solar cells that can reach power conversion efficiencies exceeding those of commercial crystalline Si devices (*1*). However, despite the rapid progress of applied research, our understanding of the basic photophysical properties of these interesting new materials is still limited. Many fundamental characteristics, including the charge carrier mobility $\mu$, photocarrier lifetime $\tau$, diffusion length $l$, and electron-hole (*e-h*) recombination coefficients $\gamma$ are not sufficiently well rationalized yet. Transient spectroscopic measurements (*2, 3*), as well as steady-state photo-Hall effect (*4*), clearly show that photocarrier lifetime and diffusion lengths in hybrid perovskites can be very long. In single crystals of methyl-ammonium lead-halides, $CH_3NH_3PbI_3$ (MAPI) and $CH_3NH_3PbBr_3$ (MAPB), $\tau$ of up to 2.7 ms and $l$ of up to 650 μm were measured, while the *e-h* recombination coefficients were found to be comparable to those in the best direct-band inorganic semiconductors like GaAs, $\gamma = 10^{-11} - 10^{-10}$ $cm^3s^{-1}$ (*4*). Such values of γ are not specifically very low from the perspective of traditional inorganic semiconductors, but they are considered to be exceptional given the inexpensive solution-based growth of perovskite materials. Radiative contribution to γ typically parallels the absorption coefficient of a material via the van Roosbroeck-Shockley's equation, and thus simultaneously low γ but high absorption coefficient in perovskites remain under debate. Although the intrinsic charge carrier mobility determined from Hall effect measurements in single crystals was found to be relatively low ($\mu = 10 - 60$ $cm^2V^{-1}s^{-1}$)(*4, 5*), its temperature dependence appears to be consistent with a band-like transport (*5*), in agreement with prior spectroscopic measurements of thin films (*2, 6, 7*). Possible ferroelectric behavior (*8*), decoration of defects and impurities with lattice polarization (*4*), large polaron formation (*4, 9*), locally indirect bandgap (*10*) and a "crystal-liquid" duality (*11*) are among the mechanisms suggested to rationalize some of the observations, including the so-called defect (trap) tolerance, which is perhaps the most attractive yet poorly understood property of lead-halide perovskites.

The proposal of an indirect bandgap has been prompted by the possibility that certain orientations of $MA^+$ dipoles may cause distortions of the surrounding $PbX_6$ octahedra, leading to a shift of the conduction band minimum (CBM) in *k*-space with respect to the valence band maximum (VBM), thus making the fundamental bandgap indirect (*10*). Another theoretical proposal involves Rashba splitting that might occur due to the strong spin-orbit coupling on Pb atom and may result in spin-allowed and spin-forbidden recombination channels, again leading to a slightly indirect character of the band gap (*12, 13*). Besides these mechanisms, it has been recently proposed that electrons and holes in hybrid perovskites might be separated in real space due to the random potential fluctuations created by disordered $MA^+$ dipoles (*14*), with the mobility of such a system best described by localization by dynamic disorder (*15*). Such a description could be compared to off-diagonal thermal disorder and transient localization models developed for organic semiconductors (*16, 17*). Localized and spatially separated carriers in this scenario can lead to a low mobility and suppressed *e-h* recombination. All these mechanisms may potentially account for very long carrier lifetimes and diffusion lengths observed in hybrid perovskites (*3, 4, 18*). The lack of consensus, however, indicates that there is an urgent need for more extensive studies focusing on highly ordered single crystals.

Here, we take advantage of the recently developed high-resolution *ac*-Hall effect measurement technique (*19, 20*), in conjunction with photoconductivity (PC) and photoluminescence (PL) spectroscopies, to be applied to several hybrid and all-inorganic perovskite single crystals. The studied compounds have different (organic or inorganic) cations, including methylammonium ($MA^+$), formamidinium ($FA^+$) and cesium ($Cs^+$). We thus compare the important transport parameters, including $\mu$, $\tau$, $l$ and $\gamma$, in this broader class of perovskites. Our photoconductivity, $\sigma_{PC}$, and photoluminescence, $I_{PL}$, measurements reveal some exciting universal patterns in *all* the studied compounds and their crystallographic phases, explained theoretically with phenomenological models proposed below.

**RESULTS AND DISCUSSION**

The community gradually comes to the realization that organic cations ($MA^+$, $FA^+$) may not play a decisive role in defining the most essential photophysical properties of hybrid perovskites. In order to verify this, we have carried out a comparative study of the hybrids (MAPB, MAPI, FAPB) and the analogous all-inorganic $CsPbBr_3$ (CPB) perovskite via photo Hall effect, photoconductivity and photoluminescence measurements performed in single crystals of each compound. Comparison of the results obtained by these complementary techniques assists in a critical assessment of microscopic mechanisms that might be governing the intriguing photophysical properties of lead-halide perovskites, including the non-trivial photoexcitation power exponents of PC and PL observed here.

**Hall effect measurements in all-inorganic CsPbBr$_3$ perovskite.**

CPB has a band gap similar to that of MAPB (*21*), and it also exhibits three structural phases (orthorhombic, tetragonal and cubic) with the two corresponding phase transitions: orthorhombic-to-tetragonal occurring at ~ 361 K and tetragonal-to-cubic occurring at ~ 403 K (*22, 23*). Representative steady-state photo-Hall effect measurements in CPB crystals carried out under a *cw* illumination with the incident photon flux $F$ in the range $7.6 \times 10^{12}$ - $3.2 \times 10^{16}$ cm$^{-2}$s$^{-1}$ are shown in Fig. 1A (for measurements details, see Methods). We'd like to emphasize that we do not see evidence of unintentional doping in this or other studied perovskites (the dark conductivities of these crystals were always much lower than photoconductivity), and thus the only asymmetry between the electron and hole populations (if any) could be that caused by trapping. Photo-Hall measurements of (undoped) band insulators, such as CPB and other lead-halide perovskites studied here, are commonly considered to yield the difference between the hole and electron mobilities, $\Delta\mu_{Hall} \equiv \mu_h - \mu_e$, which we use here as an estimate of the carrier mobility. It is important to emphasize that in CPB, $\Delta\mu_{Hall}$ is actually found to be independent of light intensity over a wide range of fluxes $F$. This light-intensity-independence is consistent with the notion that the concentrations of photogenerated electrons, *n*, and holes, *p*, are practically equal to each other for all fluxes. In the case of unequal *n* and *p*, instead of $\mu_h - \mu_e$, one would be "measuring" via Hall effect a more involved ratio $(\mu_h^2 p - \mu_e^2 n)/(\mu_h p + \mu_e n)$, that is generally expected to be $F$-dependent. The average value $\langle\Delta\mu_{Hall}\rangle$ = 9.5 cm$^2$V$^{-1}$s$^{-1}$ measured in CPB is of the same order of magnitude as $\mu$ in MAPB single crystals or large-grain-size polycrystalline MAPI films (*4*).

Figure 1B shows the corresponding photoconductivity, $\sigma_{PC}$ (blue circles), measured via 4-probe technique to address contact effects, and Hall carrier density, $n_{Hall}$ (red triangles), obtained in CPB crystal as a function of incident photon flux, $F$. It can be seen that these observables follow a power law, $\sigma_{PC}$ and $n_{Hall} \propto F^\alpha$, with the power exponent α showing a transition from the linear (α = 1) regime at low $F$ to a square-root (α = ½) regime at high $F$. We note that the incident flux corresponding to this transition, $F_T$, depends on the purity of the sample and can vary from crystal to crystal by as much as an order of magnitude. The bimolecular recombination coefficient $\gamma$, carrier lifetime $\tau$, and diffusion length $l$, can be inferred from these data using the following considerations (for more detail, see the Supplementary sec. S1 and ref. (*4*)). Under the assumption of $n \approx p$, the relationship between the bulk (3D) photoexcitation density, $G$, and photocarrier density, $n$, in steady-state measurements would be governed by the conventional rate equation:

$$\frac{dn}{dt} = \kappa G - \tau_{tr}^{-1} n - \gamma n^2 = 0. \qquad (1)$$

Here $\tau_{tr}$ is the average trap-limited carrier lifetime, and $\kappa$ is the photocarrier generation efficiency per photon (below we assume $\kappa = 1$ for simplicity). If one were to replace the 3D variables $n$ and $G$ with their projected (areal) 2D equivalents, $n_{Hall}$ and $F$, defined as $n \equiv n_{Hall}/l$ and $G \equiv F/l$, where $l$ is the carrier diffusion length that defines the effective thickness of the crystal carrying mobile photocarriers, Eq. (1) can be applied to analyze our experimental data (Supplementary sec. S1). Equation (1) illustrates the appearance of the crossover density $n_\gamma = 1/\gamma \tau_{tr}$, which separates the regimes of dominant monomolecular ($n < n_\gamma$) and bimolecular ($n > n_\gamma$) recombination. In the monomolecular regime, the quadratic term in Eq. (1) can be neglected, resulting in the power exponent $\alpha = 1$ in the $n_{Hall}(F)$ dependence. In the bimolecular recombination regime, on the other hand, the trapping term can be neglected, which leads to *e-h* recombination limited carrier lifetime $\tau = (\gamma n)^{-1}$, that is carrier density dependent, and to the power exponent $\alpha = \frac{1}{2}$, as indeed observed at higher fluxes. The crossover density in Fig. 1B corresponds to $n_{Hall} = 2.5 \times 10^{11}$ cm$^{-2}$, from which we can thus estimate the photocarrier lifetime of $\tau_{tr} \sim 100$ μs in CPB single crystals (for details of the extraction procedure see Supplementary sec. S1). The corresponding carrier diffusion length is estimated via the Einstein relationship using the experimental Hall carrier mobility $\Delta\mu_{Hall}$: $l = (\tau \cdot \Delta\mu_{Hall} \cdot k_B T/e)^{1/2} \sim 50$ μm. The *e-h* recombination coefficient would then follow from these data as $\gamma \sim 2 \times 10^{-10}$ cm$^3$s$^{-1}$ (Supplementary sec. S1). The recombination coefficient $\gamma$ thus obtained in CPB single crystals is comparable to those measured in MAPB crystals (*4*) and direct bandgap inorganic semiconductors, such as GaAs ($10^{-11}$ - $10^{-10}$ cm$^3$s$^{-1}$) (*24, 25*). The carrier lifetime in CPB appears to be somewhat shorter than $\tau_{tr}$ in MAPB, while still much longer than $\tau_{tr}$ in conventional direct-band inorganic semiconductors (~ μs) (*24, 25*). These results suggest that suppression of carrier trapping (defect tolerance) leading to remarkably long photocarrier lifetimes and diffusion lengths observed in perovskites are not solely due to organic cations, as it was previously proposed, but might be a more general property of lead-halide perovskites.

**The correlation between photoconductivity and photoluminescence.**

To further our understanding of the photo response of lead-halide perovskites, we have examined the dependence of photoconductivity and photoluminescence on the incident photon flux $F$ in the hybrid and all-inorganic perovskite single crystals. We found that both the photocurrent, $I_{PC}$, and photoluminescence power, $I_{PL}$, follow a power-law dependence on the incident photon flux: $I_{PC}, I_{PL} \propto F^{\alpha}$, where $\alpha$ is the power exponent. Moreover, we observed a clear correlation between $I_{PC}(F)$ and $I_{PL}(F)$ dependences in all the studied compounds. Figure 2A shows $I_{PC}$ and $I_{PL}$ concurrently measured in a MAPB crystal. At low incident photon fluxes, $I_{PC}$ (open squares) shows a linear ($\alpha = 1$) dependence, while $I_{PL}$ (solid squares) shows a quadratic ($\alpha = 2$) dependence on $F$. As $F$ increases, however, the power exponent $\alpha$ in photoconductivity crosses over from 1 to 1/2. The power exponent $\alpha$ in photoluminescence, on the other hand, exhibits a crossover from 2 to 3/2. The results of measurements in the inorganic CPB single crystals show a qualitatively similar behavior (Fig. 2B): the power exponent in photoconductivity changes from 1 to 1/2 (open circles), while the power exponent of photoluminescence changes from 2 to 3/2 (solid circles). Interestingly, in a much more disordered (amorphous) MAPI thin film (Fig. 2C), we have not been able to reach the crossover and only observed the first regime in the entire range of available photon fluxes, $2 \times 10^{13}$ - $4 \times 10^{17}$ cm$^{-2}$s$^{-1}$: $\alpha = 1$ in photoconductivity (open diamonds) and $\alpha = 2$ in photoluminescence (solid diamonds). In MAPI and FAPB single crystals (Fig. 2D), we measured only photoluminescence, but not photoconductivity, because the small size of the available MAPI crystals (with facets smaller than 0.1 mm) and a noticeable electric drift in FAPB crystals prevented reliable electrical measurements in these systems. Nevertheless, photoluminescence of these systems shows the consistent behavior: in MAPI, $I_{PL}(F)$ exhibits the $\alpha = 3/2$ regime (blue triangles), while $\alpha = 2$ regime is missing, because PL of MAPI is much weaker than that of MAPB, making PL spectra of this compound too noisy for a reliable extraction of power exponents at low fluxes, $F < 10^{15}$ cm$^{-2}$s$^{-1}$; FAPB, on the other hand, exhibits the same type of transition in $I_{PL}(F)$ from $\alpha = 2$ to 3/2 (green pentagons).

As we argued above, the crossover from $\alpha = 1$ to $\alpha = \frac{1}{2}$ in photoconductivity can be understood phenomenologically as being governed by the change of the dominant photocarrier decay mechanism from a trap-dominated (monomolecular) recombination to an *e-h* (bimolecular) recombination. Indeed, in the trap-dominated regime, when $\gamma n^2$ term in Eq. (1) can be neglected, a linear relationship between the steady-state photoconductivity and photoexcitation density would follow:

$$\sigma_{PC} \equiv e\mu n = e\mu\tau_{tr} \cdot G \qquad (2)$$

In the bimolecular recombination regime, when the term $\tau_{tr}^{-1} n$ in Eq. (1) is neglected, a square-root dependence of photoconductivity on the excitation density arises:

$$\sigma_{PC} = e\mu\gamma^{-1/2} \cdot G^{1/2}. \qquad (3)$$

This crossover in $\sigma_{PC}$ is indeed observed with increasing $F$ in our measurements of hybrid and all-inorganic perovskite single crystals (Fig. 2). In the MAPI thin films (Fig. 2C), the bimolecular recombination regime could not be reached, which is consistent with a much higher concentration of traps in these amorphous films.

These simple considerations, however, turn out to be insufficient when applied to results observed in photoluminescence. Indeed, assuming that photoluminescence originates from a radiative recombination of mobile electrons and holes, $I_{PL}$ would be proportional to the product of uncorrelated (free) electron and hole densities available in the sample, $I_{PL} \propto np = n^2$. Thus, in a monomolecular recombination regime, when $n$ and $\sigma_{PC} \propto F$ ($\alpha_{PC} = 1$), photoluminescence is expected to be proportional to the photoexcitation density squared, $I_{PL} \propto F^2$ ($\alpha_{PL} = 2$), while in a bimolecular recombination regime, when $n$ and $\sigma_{PC} \propto \sqrt{F}$ ($\alpha_{PC} = \frac{1}{2}$), photoluminescence is expected to follow a linear relationship, $I_{PL} \propto F$ ($\alpha_{PL} = 1$). The quadratic regime in PL is indeed observed in most of the single crystals at low incident photon fluxes and in the disordered MAPI thin films at any flux (Fig. 2). The only exception is MAPI crystals (Fig. 2D), in which we simply could not reliably measure $I_{PL}$ at sufficiently low excitation intensity to reach the quadratic regime. At high excitation fluxes, however, photoluminescence in single crystals exhibits an unusual $\alpha_{PL} = 3/2$ dependence, $I_{PL} \propto F^{3/2}$, at variance with the expectation of the linear dependence. The power exponent $\alpha_{PL} = 3/2$ observed in the bimolecular recombination regime for photoluminescence thus indicates that another microscopic mechanism governs this behavior.

**Low-*T* photoluminescence of lead-halide perovskite single crystals.**

As perovskite materials can exist in different crystallographic phases, we further access those via variable-temperature measurements. This might be helpful, for instance, in understanding whether Rashba effect has bearing on the unique optoelectronic properties of these materials. First we examine the general temperature dependence of photoluminescence spectra, because they might present a complex behavior, such as appearance of multiple peaks. Indeed, photoluminescence in perovskite thin films was reported to exhibit a band-edge shift to lower energies with cooling and emergence of additional peaks below the orthorhombic-tetragonal phase transition (*26, 27*). Figure 3 shows the temperature dependence of PL in MAPB and MAPI single crystals. With cooling, PL intensity rapidly increases, and PL peak typically shifts to longer wavelengths (red shift). As temperature is decreased further, there is a discontinuity in the PL peak position and a slight reduction in PL intensity occurring right below the orthorhombic-tetragonal phase transition in both compounds. This behavior is consistent with previous reports (*26, 27*). Note that in MAPB, photoluminescence is composed of a single band in all three phases (Fig. 3A, C), while in the orthorhombic phase of MAPI, additional peaks emerge below the phase transition at $T < 160$ K (Fig. 3B, D). PL spectra of MAPI in the orthorhombic phase were deconvoluted into four overlapping Gaussian peaks: two sharp peaks of a high intensity and two broad bands with a relatively weak emission (Supplementary Fig. S2). Here, we only analyze the two most prominent peaks: peak 1 (p1) that is present in PL starting from room temperature, and peak 2 (p2) that emerges right below the orthorhombic-tetragonal phase transition and is present in the spectra at $T < 160$ K.

Figure 4 reveals again the presence of the peculiar power exponent $\alpha_{PL} = 3/2$ in photoluminescence even at low temperature (in a different crystallographic phase). Figs. 4 A, D show photoluminescence intensity as a function of incident photon flux in the orthorhombic phase of MAPB crystals recorded at two temperatures, 80 and 120 K. The power exponents are almost exactly 3/2 ($\alpha_{PL} = 1.47$ at 80 K, and $\alpha_{PL} = 1.49$ at 120 K) in the entire investigated range of fluxes. In MAPI, we have investigated both main peaks (peaks 1 and 2 in Figs. 3 B, D), to determine the power exponent in $I_{PL}(F)$ in the orthorhombic phase. At 80 K, the power exponents are $\alpha_{PL} = 1.51$ and 1.46 for peak 1 and 2, respectively (Fig. 4B). At 120 K, the power exponents are $\alpha_{PL} = 1.41$ for both peaks (Fig. 4E). Figures 4 C, F show the PL in FAPB single crystals as a function of photoexcitation flux, also measured at 80 and 120 K. Unlike MAPB and MAPI, here the transition between $\alpha_{PL} = 2$ and 3/2 is observed at low temperatures. This appears to be consistent with the room temperature measurements shown in Fig. 2. Indeed, the transition between $\alpha_{PL} = 2$ and 3/2 in FAPB generally occurs at incident photon fluxes much higher than those in MAPB and MAPI, which can be easily seen in Fig. 2 by comparing the intercepts of the two dotted-line fits of the regimes $\alpha_{PL} = 2$ and 3/2 for these compounds. With cooling, this transition always shifts to lower fluxes (that is, the regime $\alpha_{PL} = 3/2$ becomes extended to lower $F$) in all the studied perovskites. In MAPB and MAPI at 80 and 120 K, the regime $\alpha_{PL} = 2$ is simply below the lowest accessible flux (Figs. 4, A, B, D, E). These low-$T$ measurements clearly show that the orthorhombic phase of various lead-halide perovskites (including the purely inorganic CPB) also exhibits photoluminescence scaling with flux $F$ described by power laws with the exponents 2 and 3/2, as observed in the room-temperature cubic and tetragonal phases.

Below, we discuss models that can explain the observed excitation power exponents of photoconductivity and photoluminescence in perovskites. This study helps to rule out some mechanisms and identify others as more relevant to the photophysical properties of this broad class of hybrid and inorganic perovskites. These mechanisms are based on: (a) *direct-indirect bandgap* possible in these materials, (b) *an electronic phase separation of polarons* in real space, (c) *an asymmetry in trapping rates between electrons and holes*, and finally (d) *photoexcitation flux-dependent spatial distribution of photocarriers* affected by strong *surface recombination*. We discuss advantages and disadvantages of these models in their ability to accommodate the observed signatures of system's behavior.

**(a) Direct-indirect bandgap.**

A type of bandgap called "*direct-indirect*", featuring two minima (and/or two maxima) in the conduction (and/or valence) band (*28, 29*), can be of interest for the phenomenology reported in this work. Materials with such a band structure may have a direct ($k = 0$) transition with a higher transition energy and an indirect ($k \neq 0$) transition that corresponds to the lowest (fundamental) bandgap energy, $E_g$ (Fig. 5 A). Irrespective of the microscopic origin of such a band structure, photoexcited electrons can undergo a rapid relaxation from the direct to indirect CBM, so that under a *cw* photoexcitation the equilibrium steady-state concentration of electrons in the direct CBM, $n_{dir}$, remains much smaller than that in the indirect CBM, $n_{indir}$ ($n_{dir} \ll n_{indir}$). The total concentration of photogenerated electrons is, of course, equal to that of holes: $n \equiv n_{dir} +$

$n_{indir} = p$. The photoluminescence intensity $I_{PL}$ should be mainly determined by the radiative $e$-$h$ recombination that follows the direct ($k = 0$) pathway characterized by a recombination coefficient $\gamma_{rad}$. Indeed, it is ordinarily believed that indirect transitions are much slower. Photoconductivity $\sigma_{PC}$ on the other hand should depend on both the radiative ($k = 0$) and non-radiative ($k \neq 0$, $\gamma_{non-rad}$) recombination pathways, because both of these channels contribute to the total equilibrium concentrations of mobile electrons and holes in the system. In the regime $I_{PC} \propto F^{1/2}$ and $I_{PL} \propto F^{3/2}$, occurring at sufficiently high photoexcitation intensity, the equilibrium concentration of electrons in the direct CBM is determined by the rate equation dominated by the rate $\tau_e^{-1}$ of electron relaxation to the indirect CBM: $dn_{dir}/dt = \kappa G - \tau_e^{-1} \cdot n_{dir} = 0$, resulting in $n_{dir} = \tau_e \cdot G$ (assuming $\kappa = 1$). Here, $\tau_e^{-1}$ is intended to represent a net rate in thermal equilibrium, that is a rate that includes thermal excitation back to direct CBM (also see below). On the contrary, the rate equation for the equilibrium concentration of holes is governed by their recombination with electrons in both direct and indirect CBMs: $dp/dt = G - \gamma_{rad} \cdot p n_{dir} - \gamma_{non-rad} \cdot p n_{indir} \approx G - \gamma_{non-rad} \cdot p^2 = 0$ (since $n_{dir} \ll n_{indir}$). Here, the assumption is that under a *cw* photoexcitation the indirect CBM holds most of the electrons, $n_{dir} \ll n_{indir} \approx n = p$, like a reservoir, and the dominant channel of photocarrier decay is through the indirect (and presumably slow) $e$-$h$ recombination. This gives $p = (\gamma_{non-rad})^{-1/2} \cdot G^{1/2}$ for the equilibrium concentration of holes at VBM. Thus, the photoluminescence intensity would be given by:

$$I_{PL} \propto p \cdot n_{dir} = \tau_e (\gamma_{non-rad})^{-1/2} \cdot G^{3/2}. \tag{4}$$

This behavior, governed by a direct-indirect bandgap character of the material, is consistent with our experimental observation of $\sigma_{PC} \propto F^{1/2}$ and $I_{PL} \propto F^{3/2}$ occurring at sufficiently high photoexcitation intensity. At low $F$, the photocurrent is trap dominated, and thus the regular regime with $\sigma_{PC} \propto F$ and $I_{PL} \propto F^2$ is observed. Besides being consistent with the observed power exponents, this model generally accounts well for the long photocarrier lifetimes and high absorption coefficients in perovskites. Strong absorption originates from the direct band-to-band photoexcitation, while the radiative $e$-$h$ recombination is greatly suppressed due to the presence of an "indirect reservoir" that accumulates most electrons. The main drawback of this scenario is that it requires the energy difference $\Delta E$ between the direct and indirect CBM to be substantial ($\Delta E \gg k_B T$) in order to suppress thermal backtransfer of electrons and maintain $n_{indir} \gg n_{dir}$, which would also ensure that there is no significant temperature dependence of the power exponents. The substantial $\Delta E$ however remains to be experimentally justified in these materials.

Several theoretical studies have recently suggested that long carrier lifetimes in hybrid perovskites might originate from an indirect bandgap formed via Rashba band splitting that can occur in these materials (*13, 30, 31*). Hutter *et al.* have experimentally observed features consistent with a direct-indirect bandgap character via time-resolved photoluminescence and microwave conductance measurements of polycrystalline MAPI films (*28*). The exact microscopic origin of such a band structure in perovskites still remains unclear. Indeed, in order for Rashba splitting to occur two conditions must be met: (a) a large spin-orbit coupling, and (b) an inversion symmetry of the crystal must be broken (*13, 30, 31*). The heavy Pb ion in lead-halide perovskites can result in a large spin-orbit coupling. The inversion symmetry however depends on the crystal phase. Perovskites studied here have similar structural phase transitions. For instance, MAPI undergoes a low-temperature phase transition from orthorhombic to tetragonal phase at 160 K and a high-temperature transition from tetragonal to cubic phase at 330 K (*32, 33*). In a static picture, all three phases preserve an inversion symmetry, thus suggesting that Rashba splitting should not occur (*13, 30, 31*). Nevertheless, the effect might still be possible in the hybrids due to the presence of organic dipoles (*13, 30, 31*). It has been suggested that organic dipoles can rotate quasi-freely within the perovskite lattice in the tetragonal and cubic phases (*32, 34*), and their dynamics is coupled to the motion of lead-halide octahedra, thus giving rise to an inversion symmetry breaking and Rashba effect (*13, 30, 31*). In the orthorhombic phase, on the contrary, Rashba effect should not occur, because thermal tumbling of organic dipoles stops in this low-temperature phase (*13, 30-32, 34*). Indeed, neutron scattering studies revealed that rotation of MA dipoles in MAPB completely stops below 149 K (*32, 34*). Hence, the question arises whether the direct-indirect band character that might be responsible for the observed power exponents, $\alpha_{PC} = ½$ and $\alpha_{PL} = 3/2$, actually originates from the dynamics of organic dipoles and Rashba effect.

The fact that the same power exponents are observed here both in the hybrid (MAPB, MAPI and FAPB) and in fully-inorganic (CPB) perovskites suggests that if a direct-indirect band structure is responsible for these power dependences, the origin of this band structure should not rely solely on organic cations for inversion symmetry breaking. In addition, observation of the same distinct behavior even in the low-*T* orthorhombic phase of the hybrids, where the inversion symmetry is expected to be preserved, thus suggests that either it is not the Rashba effect that is responsible for this behavior, or the inversion symmetry can be broken by local polar fluctuations of the inorganic sublattice alone, without the involvement of organic dipoles. Finally, consistently with these results, photo-Hall effect measurements of CPB show that the photocarrier lifetime and diffusion length in this compound are also quite long.

**(b) Electronic phase separation of large polarons in real space.**

An alternative microscopic picture described by the rate equations phenomenologically similar to those in the previous section and thus leading to the exponents $\alpha_{PC} = 1/2$ and $\alpha_{PL} = 3/2$ is based on the idea that electrons and holes may form large polarons that phase separate in real space (*9, 14, 35, 36*). Polaronic effects may slow down recombination by making the diffusion to a recombination site take longer, since they lead to an increased density of states of polarons as compared to free carriers, resulting in a heavier polaron effective mass (*4*). Because polaron formation is energetically favorable, its competition with thermal excitation should lead to a dynamic equilibrium between free photocarriers and large polarons under *cw* photoexcitation. Positive and negative polarons might end up in spatially separate regions, because of the opposite effect that holes and electrons exert on a highly polarizable perovskite lattice (*35-37*). In this sense, the horizontal axis in the model schematically depicted in Fig. 5A can represent a real space coordinate, *x*, and the hills (and valleys) filled with holes (and electrons) can be thought of as "puddles" in an electronically phase separated sample. Due to their physical separation, radiative recombination of such electron and hole polarons is expected to be much suppressed compared to the "direct" (non-separated) free carriers.

**(c) Electron-hole trapping asymmetry.**

A power-law dependence of photoluminescence, $I_{PL} \propto F^\alpha$, with exponent in the range $1 < \alpha < 2$ can, in principle, arise from a preferential trapping of electrons (or holes), when the traps are distributed exponentially, $n_{tr}(E) = n_{tr}^0 \cdot \exp\left(-\frac{E}{k_B T_C}\right)$, where $T_C$ is a characteristic temperature of the trap distribution, and there are some recombination states in the band gap (*38*). Under such conditions, the concentrations of photogenerated electrons and holes can be shown to follow the power law: $n, p \propto F^{T_C/(T_C+T)}$. Thus, $\sigma_{PC}$ will exhibit a power law with the exponent $T_C/(T_C + T)$, and $I_{PL}$ will have the exponent $2T_C/(T_C + T)$. Usually, $T_C > T$, and thus the power exponent of photoconductivity is in the range $\alpha_{PC} = 0.5 - 1$, while for photoluminescence $\alpha_{PL} = 1 - 2$. In one particular case, electrons can be trap dominated (that is, their concentration $n \propto F$), and holes can be dominated by a bimolecular recombination (that is, $p \propto F^{1/2}$), leading to a photoluminescence with $\alpha_{PL} = 3/2$ (Fig. 5B) (*39*). We emphasize here that the traps involved in this scenario are deep traps that quasi-permanently remove carriers from the conduction process, hence governing the carrier lifetime $\tau_{tr}$. Even though phenomenologically this picture could account for a behavior similar to that observed here, it also raises some serious questions. First, the very assumption of heavily bulk-trap dominated carriers in this model appears to be inconsistent with the notion of strongly suppressed trapping in perovskites (*4, 5, 40, 41*). More specifically, the bulk trapping model is inconsistent with other observations in our study. If photocarrier trapping with a marked asymmetry between electrons and holes was dominant, we would expect to see photo-Hall mobility to change with excitation intensity. Indeed, as more photocarriers are generated, the balance between the densities of mobile electrons, $n$, and holes, $p$, should change due to the gradual filling of trap states. According to the physics of photo-Hall effect, photo-Hall mobility can be expressed as $\Delta\mu_{Hall} = (p\mu_h^2 - n\mu_e^2)/(p\mu_h + n\mu_e)$, and when $n \neq p$ it will thus depend on these relative concentrations of mobile electrons and holes (*42*). On the contrary, our photo-Hall measurements in CPB (Fig. 1), as well as in MAPB (*4*), show that photo-Hall mobility, $\Delta\mu_{Hall}$, remains independent of the incident photon flux $F$ in a wide range (> 3 orders of magnitude). In addition, models based on traps and their distributions typically lead to a behavior that is strongly temperature dependent. Our low-$T$ measurements, however, do not reveal any qualitative differences in the behavior of power exponents $\alpha$ at low temperatures.

**(d) Variable spatial distribution of photocarriers and surface recombination.**

While the scenarios described above relied on unequal densities of electrons and holes participating in radiative recombination, the picture in this section retains the equality, $p = n$. In this sense, it represents the refinement of considerations around Eq. (1). It, however, treats the spatial profiles $n(x)$ of photocarriers as explicitly dependent on the distance $x$ from the top (illuminated) surface of the sample, $x = 0$. In our study, we use macroscopic single crystals of thicknesses 0.1 - 1 mm, which are much greater than the absorption length of the exciting light ~ 100 nm (*43-45*). It has been shown that carrier diffusion lengths $l$ in perovskites are very long, also much longer than the light absorption length. Thus photogenerated carriers diffuse deep into the crystal and occupy the volume much greater than the volume, where photogeneration takes place. Under these conditions, it is convenient to separate the effects taking place in the narrow generation region on top of the crystal from the effects taking place in the bulk. The bulk concentration $n(x)$ of charge carriers *in the steady state* then satisfies the stationary diffusion equation:

$$D\frac{d^2n}{dx^2} - \frac{n}{\tau_{tr}} - \gamma n^2 = 0, \tag{5}$$

featuring the same mono- and bimolecular bulk recombination processes as in Eq. (1). The processes taking place in the very thin light absorbing region would then be represented in the form of the boundary condition for Eq. (5):

$$-D\frac{dn}{dx}\bigg|_{x=0} = F - Sn(0), \tag{6}$$

where the diffusion current into the bulk is determined by the incident flux $F$ but reduced by the surface recombination processes (*46*). We stress that $S$ here plays the role of the effective surface recombination coefficient (also known as a surface recombination velocity) that quantifies the recombination in the whole photogeneration region. Our interest here is in thick samples, where the spatial extent of photocarriers into the bulk is limited by the recombination processes themselves rather than by the sample boundary. The model of the semi-infinite sample is then very appropriate as accompanied by the vanishing $n(x)$ and $dn/dx$ as $x$ goes to infinity. Correspondingly, one can associate the behavior of photoconductivity with that of the projected 2D carrier density, $n_{2D}$:

$$\sigma_{PC} \propto n_{2D} = \int_0^\infty n(x)dx. \tag{7a}$$

The behavior of photoluminescence, on the other hand, could then be associated with

$$I_{PL} \propto (n^2)_{2D} = \int_0^\infty n^2(x)dx. \tag{7b}$$

Under these conditions, Eq. (5) yields the exact solution for the steady-state spatial distribution $n(x)$ of photocarriers in the bulk that is determined by their concentration $n(0) \equiv n_0$ at the surface:

$$\frac{n(x)}{n_\gamma} = \frac{6A(x)}{(1-A(x))^2}, \quad A(x) = e^{-x/l}\frac{B_0-1}{B_0+1}, \quad B_0 = \left(1+\frac{2n_0}{3n_\gamma}\right)^{1/2}, \tag{8}$$

and featuring already encountered crossover density $n_\gamma = 1/\gamma\tau_{tr}$ and diffusion length $l = (D\tau_{tr})^{1/2}$. With this solution, the integral quantities (7a) and (7b) become equal to:

$$\frac{n_{2D}}{n_\gamma l} = 3(B_0 - 1) \quad \text{and} \quad \frac{(n^2)_{2D}}{n_\gamma^2 l} = \frac{3}{2}(B_0 - 1)^2(B_0 + 2). \tag{9}$$

Equations (8) and (9) clearly show that it is the ratio $n_0/n_\gamma$ that distinguishes qualitatively different limiting regimes already discussed above in the context of Eq. (1). In the trap dominated recombination regime, $n_0 \ll n_\gamma$, Eqs. (8) and (9) lead to the familiar carrier distribution $n(x) = n_0 \exp(-x/l)$ accompanied by $n_{2D} = ln_0$ and $(n^2)_{2D} = 0.5ln_0^2$ (this can also be obtained by neglecting $\gamma n^2$ term in Eq. (5)). A very different kind of solution, on the other hand, follows from Eqs. (8) and (9) for the regime dominated by the bimolecular recombination, $n_0 \gg n_\gamma$ (also obtainable by neglecting the $n/\tau_{tr}$ term in Eq. (5)). The relevant spatial behavior at $x \ll l$ then becomes a power-law decay: $n(x) = n_0/(1 + x/x_0)^2$ with the *density-dependent* length scale $x_0 = (6D/\gamma n_0)^{1/2}$, leading to $n_{2D} = (6D/\gamma)^{1/2} \cdot n_0^{1/2}$ and $(n^2)_{2D} = (2D/3\gamma)^{1/2} \cdot n_0^{3/2}$. Figure 5C clearly illustrates the density dependence of the shape of the depth profile $n(x)$ for different values of $n_0/n_\gamma$.

The analysis of the bulk Eq. (5) above determined the scaling of the photoconductivity and photoluminescence with the carrier density $n_0$ at the surface. Its dependence on the incident photon flux $F$, however, is found from the boundary condition in Eq. (6) as a result of the interplay between diffusion into the bulk and surface recombination (also shown schematically in Fig. 5C). From the exact solution of Eq. (5), this boundary condition can be conveniently rewritten as

$$n_0(S/v + B_0) = F/v, \tag{10}$$

where we defined the diffusion velocity as $v = l/\tau_{tr}$, and the interplay depends on the regime. Indeed, it is clear from Eq. (10) that in the trap-dominated regime of $n_0 \ll n_\gamma$, the surface density $n_0 = F/(S + v) \propto F$, irrespective of the magnitude of surface recombination $S$. In this regime, therefore,

$$n_{2D} \propto F \quad \text{and} \quad (n^2)_{2D} \propto F^2, \tag{11}$$

that is, precisely the scaling experimentally observed at lower fluxes $F$. The magnitude of surface recombination $S$ becomes, however, qualitatively important in the bimolecular recombination regime with $n_0 \gg n_\gamma$. Indeed, for negligible surface recombination, Eq.(10) would then yield $n_0 \propto F^{2/3}$ and, hence,

$$n_{2D} \propto F^{1/3} \quad \text{and} \quad (n^2)_{2D} \propto F. \tag{12}$$

A qualitatively different behavior takes place for substantial surface recombination. Consider the case of sufficiently large $S \gg v$, when $B_0$ in Eq. (10) can still be neglected, thus resulting in $n_0 \simeq F/S$. Note that condition $S \gg v$ seems realistic, given the low diffusion velocity $v = \frac{l}{\tau_{tr}} \approx$ 24 cm/s that can be estimated for our single crystals (*4*). The bimolecular recombination regime then would be accompanied by scaling:

$$n_{2D} \propto F^{1/2} \quad \text{and} \quad (n^2)_{2D} \propto F^{3/2}, \quad (13)$$

featuring the same power exponents for photoconductivity and photoluminescence as observed in our experiments at higher fluxes $F$.

The model crossover between the scalings in Eq. (11) and Eq. (13) upon increase of the incident photon flux $F$ is thus consistent with our experimental findings. It is peculiar that the power exponents $\alpha_{PC} = 1/3$ and $\alpha_{PL} = 1$, predicted by this model (Eq. (12)) in the bimolecular recombination regime with a negligible surface recombination (or at sufficiently high fluxes $F$), have been in fact observed in rubrene molecular crystals (*47*). The microscopic mechanism leading to this behavior in rubrene is, however, qualitatively different from this model, because it was experimentally shown to lead to an exclusively surface photoconductivity in pristine rubrene, while this model appeals to bulk photoconduction.

**CONCLUSIONS**

In summary, we have carried out a comparative study of photo-Hall effect, as well as photoconductivity and photoluminescence as functions of temperature and photoexcitation intensity in several lead-halide perovskite single crystals, including the hybrid MAPI, MAPB and FAPB and all-inorganic CPB perovskites. Photo-Hall effect in CPB reveals a moderate charge carrier mobility ~ 9.5 cm$^2$V$^{-1}$s$^{-1}$, substantial photocarrier lifetime ~ 0.1 ms and diffusion length ~ 50 μm, and a low bimolecular recombination coefficient ~ 2×10$^{-10}$ cm$^3$s$^{-1}$. Although the photocarrier lifetime appears to be an order of magnitude shorter than that in hybrid perovskites, it is nevertheless much longer than the typical values in the best direct-band inorganic semiconductors, while the other transport parameters are of the same order of magnitude as those in the hybrids. This suggests that organic dipoles in the hybrid compounds, while possibly playing a role in trap passivation, are not the main reason for the remarkably long carrier lifetimes in these materials. Our work shows that organic cations in hybrid perovskites might not be the driving force behind the possible inversion symmetry breaking and Rashba effect, if any of such exist in these materials, as evident from the measurements of the inorganic CPB and the orthorhombic phases of the hybrids at low temperature. The primary result of our work however is the observation of correlated power exponents in photoconductivity and photoluminescence, which were found to exhibit a robust crossover behavior: at relatively low illumination intensity, photoconductivity increases as a power $\alpha_{PC} = 1$ of the incident photon flux, while photoluminescence exhibits a power $\alpha_{PL} = 2$ dependence; at higher excitation intensities, photoconductivity exhibits a power $\alpha_{PC} = ½$, while photoluminescence shows a power $\alpha_{PL} = 3/2$ dependence. We proposed microscopic and phenomenological mechanisms that accommodate the observed behavior, including an electronic phase separation of electron and hole polarons in real space, a direct-indirect bandgap character, as well as the presence of substantial surface recombination.

**MATERIALS AND METHODS**

**Growth of lead halide perovskite single crystals and films.**

We prepared millimeter-size $CH_3NH_3PbBr_3$, $HC(NH_2)_2PbBr_3$ and $CsPbBr_3$ single crystals and submillimeter-size $CH_3NH_3PbI_3$ single crystals by an anti-vapor diffusion technique (*41*). For $CH_3NH_3PbI_3$ thin films, we used a two-step process. First, $PbI_2$ films were deposited on clean glass substrates by spin coating $PbI_2$ solution (462 mg of $PbI_2$ in 1 mL of dimethylformamide). Second, $PbI_2$ films were immersed in a solution of methylammonium iodide in isopropanol. After reacting with methylammonium iodide, films were rinsed with isopropanol and baked on a hot plate for 2 min at 130 °C (*48*).

**Photocurrent and photoluminescence measurements.**

Photocurrent and photoluminescence were excited with a *cw* laser with wavelength 473 nm. Photoluminescence generated in the sample was collected and imaged onto an optical fiber coupled to the Ocean Optics USB2000 spectrometer. We use long-pass filters with appropriate edge wavelengths to remove the excitation light from the detected PL signal. In photocurrent measurements, electrical contacts are prepared by sputtering of high purity metals (Ti/Pt) at the crystal's surface through a shadow mask. The contacts are formed in a coplanar configuration with inter-contact separation ranging from 0.5 to 1.5 mm. Photocurrent was recorded by using Keithley source-meter (K-2400) and electrometer (K-6514) controlled by a LabVIEW program.

All measurements reported here are carried out in high vacuum (~ $10^{-5}$ Torr). A closed cycle cryostat (Advanced Research Systems) was used for temperature variable measurements.

**Hall effect measurements.**

*ac* Hall-effect measurements were performed at room temperature in air. We use an SR-830 lock-in amplifier (Stanford Research) and a low-noise voltage preamplifier 1201 (DL instruments) for Hall voltage measurements, and a *dc* current source K-6221 (Keithley) for current excitation. Photocarriers were excited by uniformly illuminating the sample's surface with a monochromatic light ($\lambda$ = 465 nm) emitted by a calibrated high-power blue LED. An *ac* magnetic field of an rms magnitude 0.23 T and frequency in the range 0.5 - 3 Hz was generated by rotating an assembly of permanent Nd magnets. The magnetic field was applied perpendicular to the sample's surface, while a *dc* current flown in the longitudinal direction between source and drain contacts. Longitudinal sample's conductivity or photo-conductivity was measured by four-probe technique to correct for contact resistance. 4-probe longitudinal voltage was measured *in-situ* by using a Keithley electrometer (K-6514). The photo-Hall mobility is calculated as: $\Delta\mu_{Hall} \equiv \mu_h - \mu_e = (V_{Hall}/W) \cdot (d_{4p}/V_{4p}) \cdot B^{-1}$, where $d_{4p}$ is the distance between the voltage probes in the four-probe contact geometry, $W$ is the channel width, and $V_{4p}$ is the longitudinal voltage drop between the voltage probes (*4, 42*).

**ACKNOWLEDGMENTS**

H.T.Y. and V.P. thank the National Science Foundation for the financial support of this work under the grant DMR-1506609 and the Institute for Advanced Materials and Devices for Nanotechnology (IAMDN) of Rutgers University for providing necessary facilities. X.-Y.Z. acknowledges support by the US Department of Energy, Office of Science - Basic Energy Sciences, Grant ER46980, Y.N.G. is grateful for support from the Department of Energy, Office of Basic Energy Science (DOE/OBES) grant DE-SC0010697. We thank Dr. Jinghui He of the School of Chemistry, Chemical Engineering and Materials Science, Soochow University, China for his help with the growth of FAPB crystals.

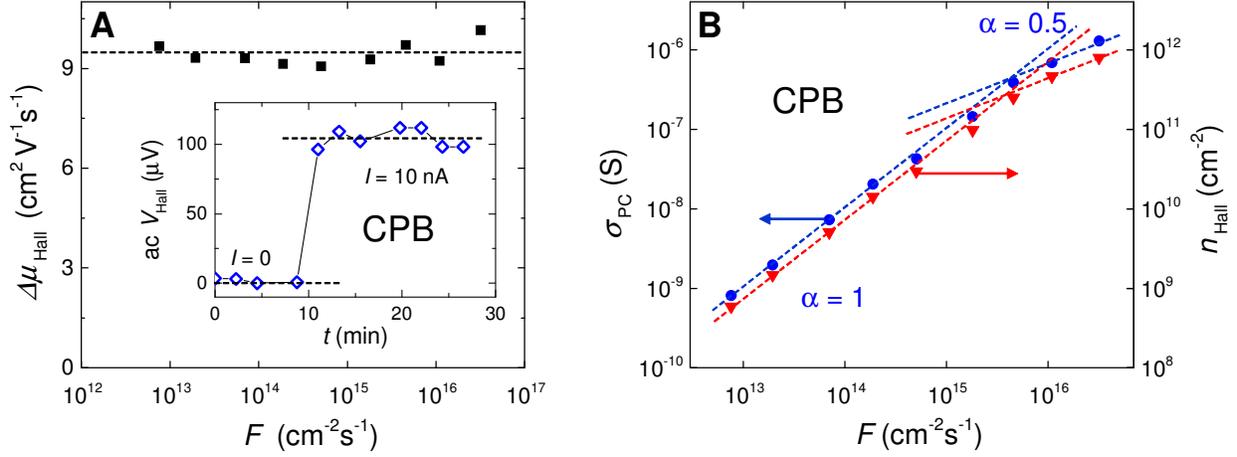

**Figure 1. Photo-Hall effect measurements in all-inorganic CsPbBr$_3$ single crystals.** (**A**) Photo-Hall mobility, $\Delta\mu_{\text{Hall}} \equiv \mu_h - \mu_e$, measured as a function of incident photon flux, $F$, varied over several orders of magnitude (squares). A nearly constant value of $9.5 \pm 0.6$ cm$^2$V$^{-1}$s$^{-1}$ is observed (dashed line represents the average). The inset shows a representative $ac$-$V_{\text{Hall}}$ signal (diamonds) recorded at $ac$-$B$ field of rms magnitude 0.23 T at 0.6 Hz under incident photon flux of $7\times10^{13}$ cm$^{-2}$s$^{-1}$ ($\lambda$ = 465 nm). Two levels of $V_{\text{Hall}}$, corresponding to zero and non-zero (10 nA) longitudinal $dc$ excitation current, are clearly observed. For details of the technique see Ref. (*19*). (**B**) 4-probe photoconductivity, $\sigma_{\text{PC}}$ (blue circles), and Hall carrier density, $n_{\text{Hall}}$ (red triangles), measured as a function of incident photon flux, $F$. Dashed lines are the power-law fits with exponents $\alpha$ = 1 and ½ (as indicated). A transition from monomolecular to bimolecular recombination regimes is observed.

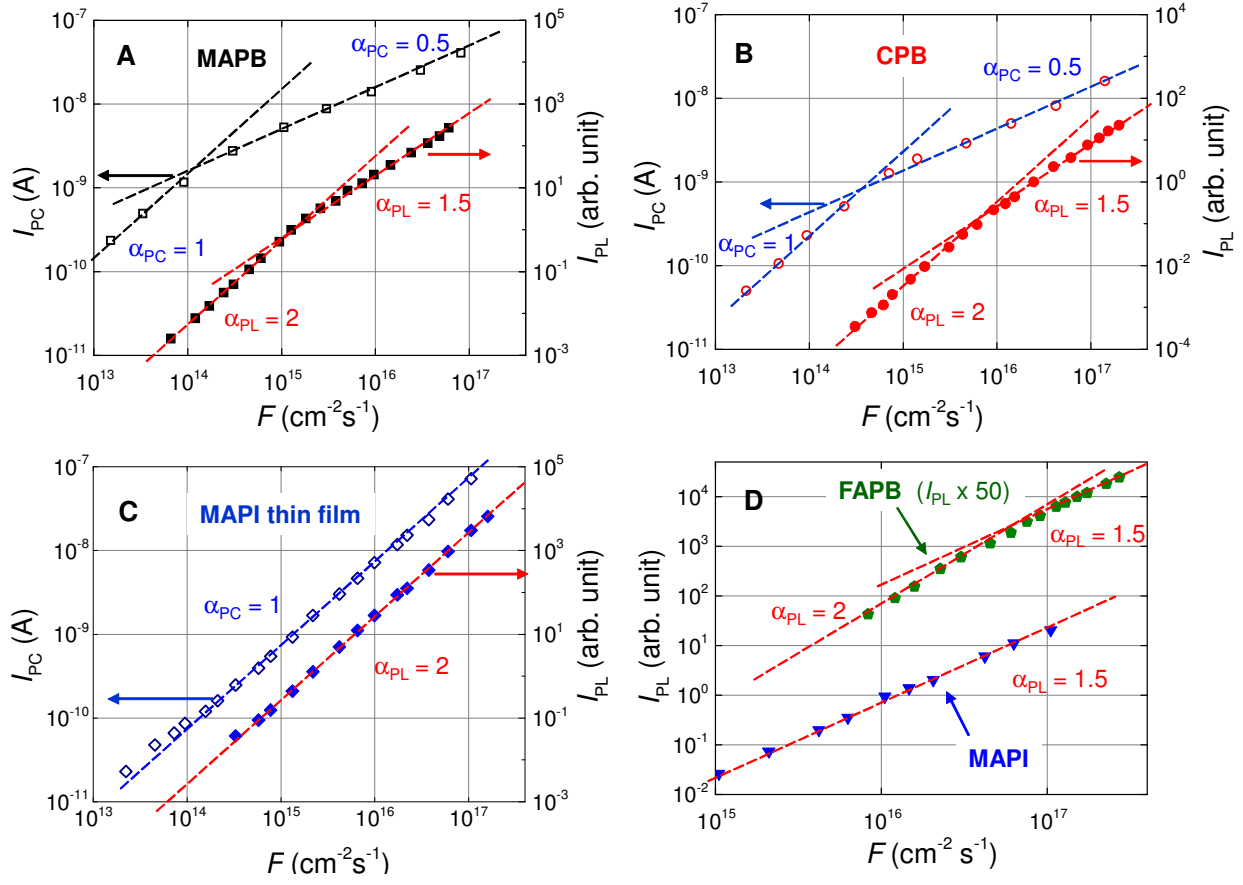

**Figure 2. Photocurrent, $I_{PC}$, and photoluminescence intensity, $I_{PL}$, in several lead-halide perovskites as a function of incident photon flux, $F$.** (**A**) $I_{PC}$ (open squares) and $I_{PL}$ (solid squares) measured in MAPB single crystals. (**B**) $I_{PC}$ (open circles) and $I_{PL}$ (solid circles) measured in CPB single crystals. (**C**) $I_{PC}$ (open diamonds) and $I_{PL}$ (solid diamonds) measured in a disordered (amorphous) MAPI thin film. (**D**) $I_{PL}$ measured in FAPB (green pentagons) and MAPI (blue triangles) single crystals. Dotted lines in all panels are the power-law fits with power exponents $\alpha$ = 1/2, 1, 3/2 and 2 (as indicated). Measurements were carried out at room temperature.

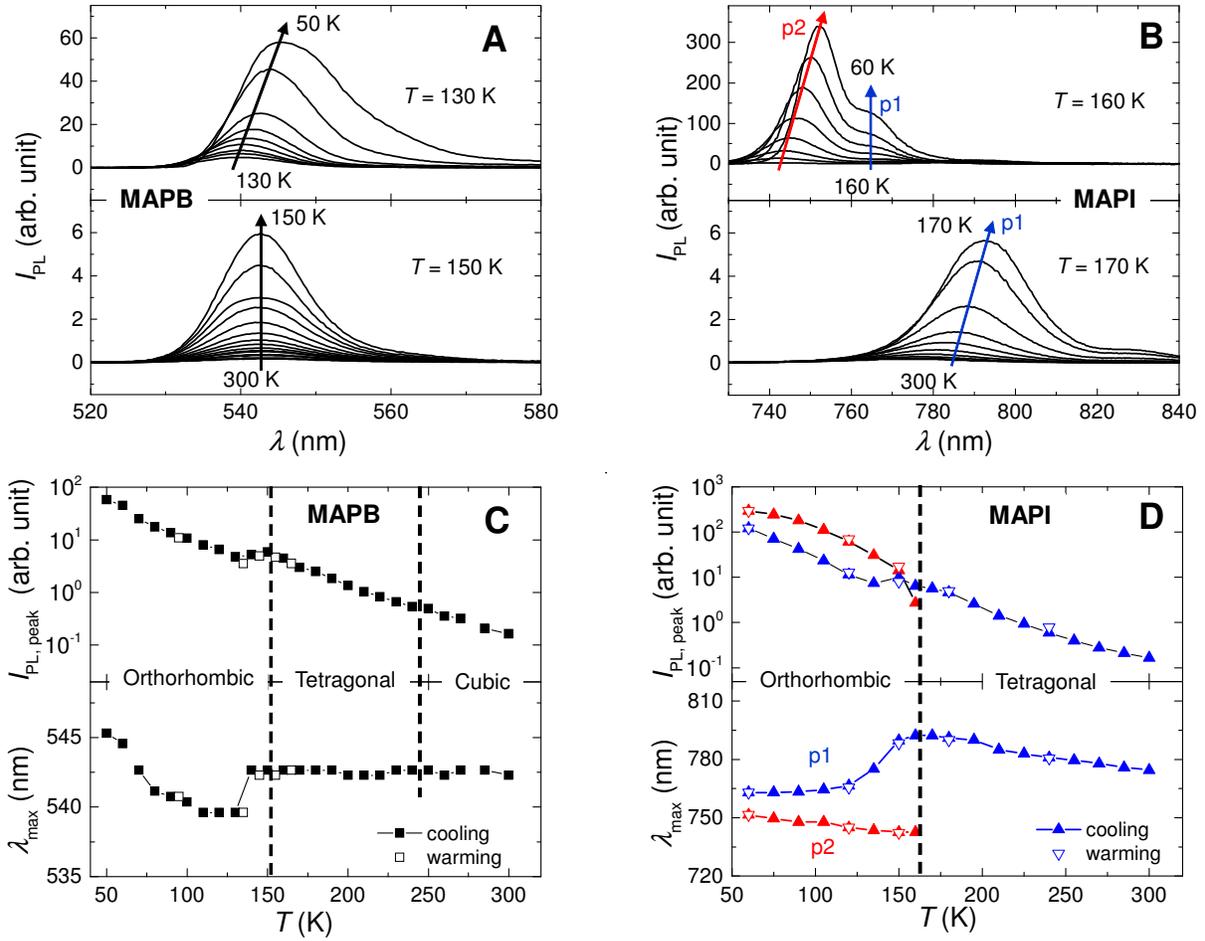

**Figure 3. Evolution of photoluminescence of MAPB and MAPI single crystals with temperature.** (**A**) PL spectra recorded at different temperatures in the orthorhombic (upper panel), as well as tetragonal and cubic (lower panel) phases of MAPB crystals (photoexcitation: $\lambda = 473$ nm, $F = 4 \times 10^{16}$ cm$^{-2}$s$^{-1}$). Arrows indicate the change in $T$ in 5-15 K steps. (**B**) PL spectra recorded at different temperatures in the orthorhombic (upper panel), as well as tetragonal and cubic (lower panel) phases of MAPI crystals (photoexcitation: $\lambda = 473$ nm, $F = 1.5 \times 10^{16}$ cm$^{-2}$s$^{-1}$). Two PL peaks, denoted p1 and p2, are identified by multiple peak deconvolution (Supplementary Fig. S2). (**C**) PL intensity (upper panel, semi-log scale) and peak position (lower panel) in MAPB crystals as a function of temperature in all three phases. (**D**) PL intensity (upper panel, semi-log scale) and peak positions (lower panel) in MAPI crystals as a function of temperature (two PL peaks, p1 and p2, are analyzed). Open and solid symbols correspond to the data collected on warming and cooling, respectively. Vertical dashed lines mark the approximate phase transition temperatures.

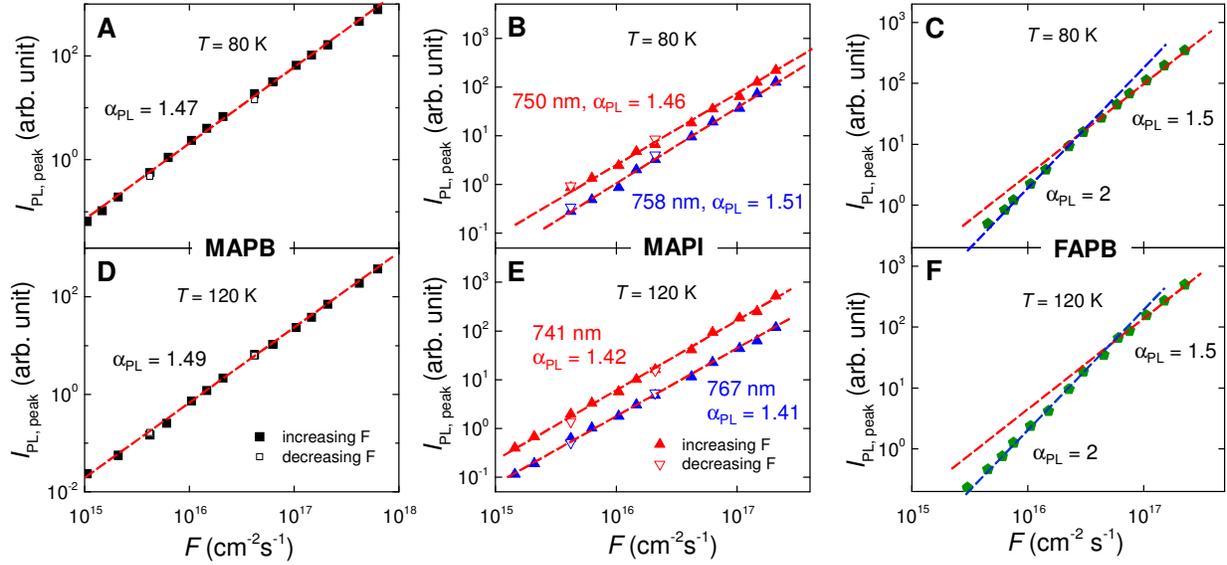

**Figure 4. Dependence of photoluminescence intensity on the incident photon flux, $I_{PL}(F)$, in the orthorhombic phase of MAPB, MAPI and FAPB single crystals.** The PL spectra were recorded under a *cw* excitation ($\lambda$ = 473 nm). Note double-log scale in all panels. (**A**) MAPB, (**B**) MAPI and (**C**) FAPB single crystals measured at 80 K, and (**D**) MAPB, (**E**) MAPI and (**F**) FAPB single crystals measured at 120 K. All of these systems are in the orthorhombic phase at these temperatures. A single PL peak of MAPB and FAPB and the two most prominent PL peaks of MAPI (labeled p1 and p2 in Fig. 3D) were traced. Solid and open symbols represent measurements on increasing and decreasing *F*, respectively. Dashed lines are the power-law fits with the power exponents $\alpha$ (indicated). It is clear that $\alpha$ is very close to the ideal 2 or (mostly) 3/2 in all these systems. In FAPB, the transition between $\alpha$ = 2 and 3/2 is seen even at these low temperatures, while MAPB and MAPI only exhibit $\alpha$ = 3/2 in the studied range of photoexcitation fluxes. There is no hysteresis in $I_{PL}(F)$ in these measurements.

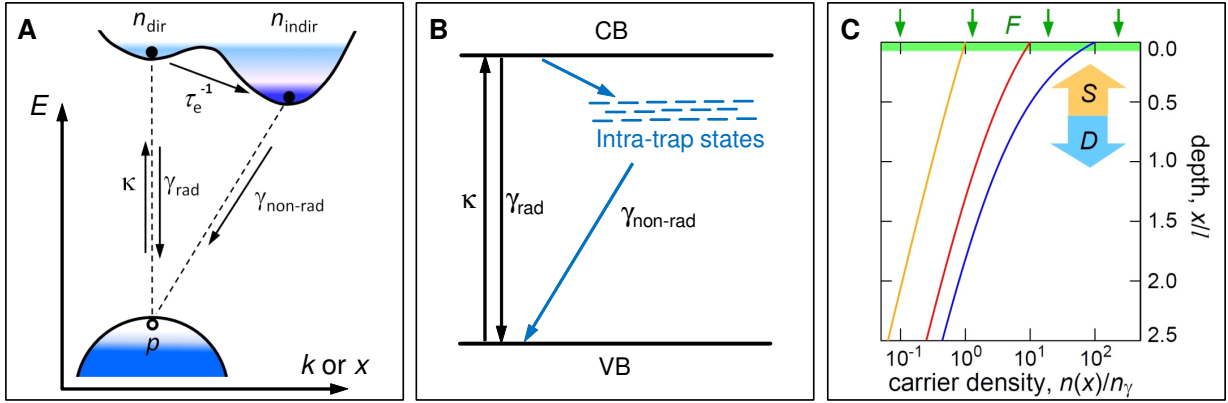

**Figure 5. Phenomenological models that can account for the observed power exponents in PC and PL of lead-halide perovskites.** (A) Direct-indirect bandgap structure with the relevant processes indicated by the arrows, and $p$ and $n = n_{dir} + n_{indir}$ denoting the densities of mobile holes and electrons, respectively. (B) Preferential trapping of one type of charge carriers (here, electrons). (C) Illustration of the density-dependent spatial profiles $n(x)$ of the photoexcited charge carrier density calculated with Eq. (8) for three different values of the surface carrier concentration $n_0$. The value of the latter is established by the incident photon flux $F$ as a result of the interplay between the diffusion into the bulk (coefficient $D$) and surface recombination (coefficient $S$), as per Eq. (10). In this model, it is assumed that the incident photon flux $F$ is fully absorbed at the very surface of the sample (at $x/l = 0$), and the sample's thickness is much greater than the carrier diffusion length.